\documentclass[11pt]{article}
\usepackage{amsfonts}
\usepackage{graphicx}
\usepackage{epstopdf}
\usepackage{epsfig}
\usepackage{amssymb}
\usepackage{setspace}
\usepackage{caption}
\usepackage{amsfonts}
\usepackage{color}
\usepackage{amsmath}
\usepackage{float}
\usepackage{comment}
\newcommand {\be}{\begin{equation}}
\newcommand {\ee}{\end{equation}}
 \newcommand {\bea}{\begin{array}}
 
 \newcommand {\eea}{\end{array}}

\textwidth=6.5in \hoffset=-.75in \voffset=-1in \numberwithin{equation}{section}
\numberwithin{figure}{section}

\begin{document}

\begin{titlepage}
\vspace{1cm} 
\begin{center}
{\Large \bf {Hidden conformal symmetry for the accelerating Kerr black holes}}\\
\end{center}
\vspace{2cm}
\begin{center}
\renewcommand{\thefootnote}{\fnsymbol{footnote}}
Haryanto M. Siahaan{\footnote{haryanto.siahaan@gmail.com}}\\~~\\
MTA Lend\"{u}let Holographic QFT Group, Wigner Research Centre for Physics,\\
Konkoly-Thege Mikl\'{o}s u. 29-33, 1121 Budapest, Hungary\\and\\
Center for Theoretical Physics, Physics Department, Parahyangan Catholic University,\\
Jalan Ciumbuleuit 94, Bandung 40141, Indonesia
\renewcommand{\thefootnote}{\arabic{footnote}}
\end{center}

\begin{abstract}
According to Kerr/CFT correspondence, some aspects of extremal and non-extremal Kerr black holes could be explained by using a 2D CFT. This paper is devoted to show how Kerr/CFT correspondence can work for the slowly accelerating Kerr black holes, in both extremal and non-extremal cases. The equation of motion for low frequency scalar probes in the near region possesses the ``hidden'' conformal symmetry, similar to that of the non-accelerating version. Assuming a particular value of associated central charge, Cardy formula in 2D CFT can recover the Bekenstein-Hawking entropy for the black hole under consideration. Moreover, agreements in the scattering process between gravitational and 2D CFT dual suggest the existence of slowly accelerating Kerr/CFT duality.
\end{abstract}
\end{titlepage}\onecolumn 
\bigskip 

\section{Introduction}
\label{sec:intro}

Black holes are among the most compelling objects in our universe. The spacetime with strong gravity around black holes provides an arena to explore aspects of a gravitational theory, Einsteinian or beyond. Many exact black hole solutions are available in literature, coming from various theories containing gravity. One of the most famous is the Kerr solution, i.e. the asymptotically flat rotating metric which solves the vacuum Einstein equations \cite{Kerr:1963ud,Hawking:1971vc}. A Kerr black hole is characterized by its mass and angular momentum, to which extensive studies have been done. Some of the important examples are, among others, the electromagnetic energy extraction \cite{Blandford:1977ds}, the astronomical observation \cite{McClintock:2006xd}, uniqueness of the solution \cite{Robinson:1975bv}, and the separability of test scalar field's equation in Kerr background \cite{Teukolsky:1973ha}.

The advent of AdS/CFT holography \cite{Maldacena:1997re,Gubser:1998bc,Witten:1998qj,Aharony:1999ti} changes the direction of physics researches in a quite broad spectrum. Inspired by this holography, the authors of \cite{Guica:2008mu} proposed that the physics of extremal Kerr black holes can be understood using a 2D CFT as the dual description. The $SL\left( {2,R} \right)_L  \times U\left( 1 \right)$ symmetry that is possessed by the near horizon of an extremal Kerr black hole suggests the possibility to adopt the classic work on AdS$_3$/CFT$_2$ by Brown and Henneaux \cite{Brown:1986nw} in establishing the holography between extremal Kerr black holes and 2D CFT. This is exactly what Guica et. al. showed in \cite{Guica:2008mu}, and by Carlip using an alternative way \cite{Carlip:2011ax}, which has been extended to many cases of rotating black holes since then \cite{Compere:2012jk,Siahaan:2015xia}. 

However, when the Kerr black holes are not extreme, a different approach has to be taken in showing the holography. Instead of looking for the conformal symmetry in its near horizon geometry, the $SL\left( {2,R} \right)_L  \times SL\left( {2,R} \right)_R$ symmetry is found to be hidden in the near region equation of motion for a low frequency scalar probe \cite{Castro:2010fd}. Assuming the central charge in non-extremal case is smoothly connected to that of the extreme one, the non-chiral Cardy formula of 2D CFT recovers the Bekenstein-Hawking entropy for a non-extremal Kerr black hole. The temperature in each left and right mover dual 2D CFT theory are found by matching the corresponding squared Casimir operator to the Laplacian appearing in the radial equation for the scalar probe. In addition to this matching entropy, the holography can be seen from the scattering process where the gravitational result for absorption cross section is in agreement to that of 2D CFT calculation. It seems that this hidden conformal symmetry is a property of any rotating black hole, and moreover the authors of \cite{Chen:2010ik} argued that this ``hidden'' conformal symmetry is not just a special feature in the wave equation but intrinsically of the spacetime. Investigations along this line have been performed for many type of black holes \cite{Ghezelbash:2012qn,Chen:2010as,Chen:2010fr,Chen:2010ywa,Chen:2011kt}. Interestingly, in the style of hidden conformal symmetry \cite{Franzin:2011wi}, the author of \cite{Majhi:2015tpa} managed to reveal the hidden conformal symmetry that is possessed by a class of time dependent spacetime. Moreover, very recently, the long standing question on how to show the hidden conformal symmetry for higher spin probes living in a stationary spacetime is answered in \cite{Shi:2018lpw}.

In a vacuum Einstein system, there also exists an exact solution describing a pair of black holes constantly accelerating away each other. It is known by the name of $C$-metric \cite{Kinnersley:1970zw,Dias:2002mi,Plebanski:1976gy,Griffiths:2005qp,Griffiths:2009dfa,Hong:2003gx}, and the rotating version expressed in Boyer-Lindquist type of coordinate is given by Hong and Teo in \cite{Hong:2004dm}. Normally, in the case of accelerating black holes, the existing conical singularities take different values depending on which axis is under consideration, i.e. $\theta =0$ or $\theta =\pi$. Without introducing external electromagnetic field, only one of these nodal singularities that can be removed by rescaling angular coordinate \cite{Astorino:2016xiy}. The other conic singularity which still exists after rescaling the coordinate is then interpreted as the source of black hole acceleration \cite{Griffiths:2009dfa}. Note that the notion of ``accelerating black holes`` could mislead the reader due to the absent of traditional black holes\footnote{A physical singularity covered by even horizon(s).} in the spacetime under consideration, due to the existence of the cosmic singularity along its rotational axis. Nevertheless, the use of black hole term helps intuitively to understand that we are dealing with a mass in the spacetime, and recognizing some analogous properties such as area and temperature \cite{Appels:2016uha,Astorino:2016ybm,Appels:2017xoe}. Alternatively, the reader can view this work as the generalization of Kerr/CFT relation \cite{Castro:2010fd} to the case of rotating $C$-metric. Since this accelerating black hole exactly solves the vacuum Einstein equation, there should be a thermodynamical relation that associates to it, in analogy to the one of non-accelerating black holes. These interesting investigations have been performed in \cite{Appels:2016uha,Astorino:2016ybm} for cosmic string with non varying tension, and in \cite{Appels:2017xoe} where the it can vary. 

Due to these interesting features, investigating the Kerr/CFT holography for an accelerating Kerr black hole should be an interesting task. In \cite{Astorino:2016xiy}, the author showed the validity of Kerr/CFT conjecture \cite{Guica:2008mu} in the case of extremal accelerating Kerr black hole. Moreover, the author of \cite{Astorino:2016xiy} also hints the existence of non-extremal version of Kerr/CFT duality \cite{Castro:2010fd} for accelerating black holes, motivated by the separability of Klein-Gordon equation for scalar probe \cite{Bini:2014kga}. This is exactly what we pursue in this paper, showing the hidden conformal symmetry that belongs to a slowly accelerating Kerr black hole, and establish the Kerr/CFT duality from it. In addition to the non-extremal case, the extremal version of hidden conformal symmetry \cite{Chen:2010fr} is also investigated for this black hole.

The organization of this paper is as follows. In the next section we review some properties of an accelerating Kerr black hole. In section \ref{sec.KG}, we discuss the Klein-Gordon equation for a scalar probe, and obtain its low frequency limits in the near region. In subsequent section, we reveal the conformal symmetry that is hidden in the radial equation for both non-extremal and extremal cases. Dual calculation for entropy and scattering absorption cross section are given in section \ref{sec.holography}. Finally, we give discussions and conclusions in the last section. We also have appendices where the alternative calculation for central charge and Hawking temperature are given.

\section{Accelerating Kerr black spacetime}\label{sec.reviewACC}

Accelerating black holes are interesting spacetime solution in general relativity. Its static version was found by Levi-Civita in 1918, and then by Weyl in 1919, and subsequently rediscovered by many authors later on \cite{Griffiths:2009dfa}. This black hole solution can exist in the Einstein-Maxwell theory, even with the presence of cosmological constant. The common interpretation for this solution is a pair of black holes which are constantly accelerating away each other, where the existing nodal singularity provides the acceleration. In last couple years, discussions on thermodynamical aspects of this black holes is among interests of several authors \cite{Appels:2016uha,Appels:2017xoe,Astorino:2016ybm,Astorino:2016xiy}. Some interesting aspects which are raised in these works cover the conserved quantities definitions for such black holes and connection between the change of cosmic string tension to the black hole mechanics formula. It appears that accelerating black holes have some non-trivial features which do not exist in the non-accelerating version. In this section we briefly review some aspects of an accelerating black hole, which are useful in the proceeding discussions. 

One way to express the accelerating Kerr black hole is using Hong and Teo set up \cite{Hong:2004dm}, which directly reduces to the Boyer-Lindquist form of the Kerr black hole. In Hong-Teo style, the line element describing rotating and accelerating black holes is
\be \label{metric.acc.Kerr.ori}
ds^2  = \Omega ^{ - 2} \left[ { - \frac{Q}{{\rho ^2 }}\left( {dt - a\sin ^2 \theta d\phi } \right)^2  + \rho ^2 \left( {\frac{{dr^2 }}{Q} + \frac{{d\theta ^2 }}{P}} \right) + \frac{{Pa^2 \sin ^2 \theta }}{{\rho ^2 }}\left( {dt - \frac{{r^2  + a^2 }}{a}d\phi } \right)^2 } \right]\,,
\ee 
where
\be
\Omega  = 1 - \alpha r\cos \theta \,,
\ee
\be
\rho ^2  = r^2  + a^2 \cos ^2 \theta \,,
\ee
\be
P = 1 - 2\alpha M\cos \theta  + \alpha ^2 a^2 \cos ^2 \theta \,,
\ee
\be
Q = \left( {r^2  - 2Mr + a^2 } \right)\left( {1 - \alpha ^2 r^2 } \right)\,.
\ee
This metric solves the vacuum Einstein equation $R_{\mu\nu}=0$ and represents the spacetime containing a pair of constantly accelerated and rotating black holes. The parameters appearing in this line element are as follows, $M$ is black hole mass, $a$ is the rotational parameter, and $\alpha$ is the acceleration of black holes. Setting $\alpha =0$ yields the famous asymptotically flat Kerr spacetime, and the condition $a =0$ gives the $C$-metric.

To verify that this solution really describes a pair of constantly accelerating objects, one can take the non-rotating and Minkowski limits of such solution, i.e. taking $a\to 0$ and $M\to 0$ respectively. Then consider any worldline $x^\mu \left(\tau\right) = \left(t\left(\tau\right),r_0,\theta_0,\phi_0\right)$, where $r_0$, $\theta_0$, and $\phi_0$ are some fixed coordinates. The corresponding four-velocity to this worldline is $u^\mu = \left({\dot t},0,0,0\right)$. The magnitude of acceleration can be found by calculating the scalar product of acceleration vector $a^\nu   = u^\mu  \nabla _\mu  u^\nu $, which turns out to be $a^\mu  a_\mu   =  - \alpha ^2 $.

Now let us turn to the nodal deficit and excess that an accelerating Kerr spacetime contains. Near the north and south poles, i.e. at $\theta = 0$ and $\theta = \pi$ respectively, one round of a circle is not exactly $2\pi$ as in the spacetime without nodal singularity. The deficit/excess angles near these poles are
\be \label{Cplus}
\mathop {\lim }\limits_{\theta  \to 0} \frac{{2\pi }}{{\sin \theta }}\sqrt {\frac{{g_{\phi \phi } }}{{g_{\theta \theta } }}}  = 2\pi \Xi_+
\ee
and
\be \label{Cmin}
\mathop {\lim }\limits_{\theta  \to {\pi}} \frac{{2\pi }}{{\sin \theta }}\sqrt {\frac{{g_{\phi \phi } }}{{g_{\theta \theta } }}}  = 2\pi \Xi_-
\ee
where $\Xi_\pm = {1 \mp 2\alpha m + \alpha ^2 a^2 }$. These singularities cannot be removed simultaneously, unless some external electromagnetic fields are introduced \cite{Astorino:2016xiy}. Nevertheless, in this article we do not consider the spacetime with external fields, hence a nodal singularity still exists after fixing one of them.

In this paper we choose to consider the accelerating black hole with nodal singularity at $\theta =0$, by fixing the excess angle near $\theta = \pi$. This can be done by rescaling $\phi \to \phi /\Xi_- $, and from now on the subscript ``-'' will be suppressed for simplicity\footnote{Unless it is needed to be restored to make distinction with $\Xi_+$.}. The deficit angle at the $\theta =0$ axis is interpreted to be a semi-infinite cosmic string connecting the $r=0$ and a point at infinity, and this string which pulls the back hole to have an acceleration. The resulting line element after performing this rescaling reads
\be \label{metric.acc.Kerr.fixed.nodal}
ds^2  = \Omega ^{ - 2} \left[ { - \frac{Q}{{\rho ^2 }}\left( {dt - \frac{a\sin ^2 \theta}{\Xi}  d\phi } \right)^2  + \rho ^2 \left( {\frac{{dr^2 }}{Q} + \frac{{d\theta ^2 }}{P}} \right) + \frac{{Pa^2 \sin ^2 \theta }}{{\rho ^2 }}\left( {dt - \frac{{r^2  + a^2 }}{\Xi a}d\phi } \right)^2 } \right]
\ee 
where the deficit angle near the $\theta=0$ axis is
\be 
\Delta \phi  = \frac{{8\pi \alpha M}}{{1 + 2\alpha M + \alpha ^2 a^2 }}\,.
\ee 

To end this section, let us provide some properties of an accelerating Kerr black hole, which are useful in the subsequent discussions. The area of black hole is given by
\be 
A_{BH}  = \int\limits_0^{2\pi } {d\phi } \int\limits_0^\pi  {d\theta \left. {\sqrt {g_{\phi \phi } g_{\theta \theta } } } \right|_{r = r_ +  } }  = \frac{{8\pi M r_+}}{\Xi \left({1 - \alpha ^2 r_ + ^2 }\right)}\,,
\ee
the angular velocity at horizon reads
\be 
\Omega _H  = \left. { - \frac{{g_{t\phi } }}{{g_{\phi \phi } }}} \right|_{r = r_ +  }  = \frac{a \Xi}{{2Mr_+}}\,,
\ee 
and the conserved angular momentum is found to be
\be 
J = \frac{1}{{16\pi }}\int_S {\nabla ^\mu  \zeta ^\nu  d\Sigma _{\mu \nu } }  = \frac{Ma}{\Xi^2}
\ee
where $\zeta ^\mu $ is the axial Killing vector and $d\Sigma _{\mu \nu }  =  - 2n_{[\alpha } \sigma _{\beta ]} \sqrt {g_{\theta \theta } g_{\phi \phi } } d\theta d\phi $. Vectors $n^\alpha   = \left[ {1,0,0,0} \right]$ and $\sigma^\alpha   = \left[ {0,0,0,1} \right]$ are orthonormal each other, and normal to the surface $S$. The entropy of accelerating black hole is given by \cite{Appels:2016uha,Astorino:2016xiy}
\be
S_{BH}  = \frac{{A_{BH} }}{4} = \frac{{2\pi Mr_ +  }}{{\Xi \left( {1 - \alpha ^2 r_ + ^2 } \right)}}\,,
\ee 
and the Hawking temperature reads\footnote{An alternative derivation of this Hawking temperature using tunneling method is given in appendix \ref{sec.app2}.}
\be \label{THaccS}
T_H  = \frac{{\left( {r_ +   - r_ -  } \right)\left( {1 - \alpha ^2 r_ + ^2 } \right)}}{{8\pi Mr_ +  }}\,.
\ee

\section{Massless scalar perturbations in accelerating Kerr background}\label{sec.KG}

Now in this section, let us first review the equation of motion for scalar probes in an accelerating Kerr background, and then obtain the near region and low frequency limits of that equation for a slowly accelerating case. It is well known that a set of Teukolsky-like master equations can be obtained for such spacetime \cite{Dudley:1977zz,Dudley:1978vd}. In an interesting work \cite{Bini:2014kga}, the authors discussed massless and neutral probes in the background of accelerating Kerr black holes. They employ the Newman-Penrose formalism to obtain a master equation of motion for test particles with spins $s\le 2$. The result is analogous the Teukolsky master equation for Kerr background \cite{Teukolsky:1973ha}. 

Recall that, using the ansatz for a scalar probe
\be \label{Ansatz.sep.Teu}
\Phi  = e^{ - i\left( {\omega t - m\phi } \right)} R\left( r \right)S\left( \theta  \right)\,,
\ee
Teukolsky was the first one who show that the equation of motion for a free scalar field $\Phi$ in Kerr background is separable between its angular and radial dependent parts. Interestingly, Bini et. al. \cite{Bini:2014kga} also managed to show the separability of Klein-Gordon equation in an accelerating Kerr background by using a slightly more general ansatz compared to (\ref{Ansatz.sep.Teu}). This separability which hints the existence of hidden conformal symmetry in the accelerating Kerr background, as pointed out in \cite{Astorino:2016xiy}. Based on this hints, we thoroughly investigate the hidden conformal symmetry for an accelerating Kerr black hole, in both non-extremal and extremal cases. Another interesting works on the test fields in an accelerating black hole background can be found in \cite{Bini:2015xpa,Kofron:2015gli,Kofron:2016dyk}, where further discussions on the solution's separability and also Dirac field perturbation are presented. 

In this paper, we limit our discussions for the neutral and massless scalar test field only. The corresponding equation of motion related to this field, in the background of rotating and accelerating black holes, is given by
\be 
\nabla _\mu  \nabla ^\mu  \Phi  = 0\,.
\ee 
The last expression is exactly the case $s=0$ of the master equation in \cite{Bini:2014kga}. Slightly different to the separation ansatz (\ref{Ansatz.sep.Teu}), for the accelerating Kerr black hole we should use \cite{Bini:2014kga}
\be \label{KG1}
\Phi  = \left(1-\alpha r \cos\theta\right) ~ e^{ - i\left( {\omega t - m\phi } \right)} R\left( r \right)S\left( \theta  \right)\,,
\ee 
i.e. the ansatz (\ref{Ansatz.sep.Teu}) scaled by $\Omega$. In the case of non-accelerating, i.e. $\alpha = 0$, this anstaz is just the one employed in generic Kerr background (\ref{Ansatz.sep.Teu}). This separation method allows the field equation (\ref{KG1}) to be separated into the radial and angular equations, which read \cite{Bini:2014kga}
\be\label{eq.radial}
\left[ {\partial _r Q\partial _r  - 2r\alpha ^2 \left( {r - M} \right) + \frac{{\left( {\left( {r^2  + a^2 } \right)\omega  - am\Xi} \right)^2 }}{Q} - {\tilde \lambda} } \right]R\left( r \right) = 0 \,,
\ee
and
\be\label{eq.ang} 
\left[ {\frac{1}{{\sin \theta }}\partial _\theta  \sin \theta \partial _\theta   - \frac{{ m^2 \Xi^2 \cot ^2 \theta  - a^2 \omega^2 \cos ^2 \theta  + \left( {a\omega  - m\Xi} \right)^2  - L\left( \theta  \right)}}{{P^2 }} + \frac{{\tilde \lambda} }{P}} \right]\sqrt P S\left( \theta  \right) = 0 \,,
\ee
respectively. In the equation above, $L\left( \theta  \right) = \alpha ^2 \sin ^2 \theta \left( {M^2  - a^2 } \right)$, and ${\tilde \lambda}$ is the separation constant.

Since we are investigating the hidden conformal symmetry for a slowly accelerating version of the black hole discussed in \cite{Castro:2010fd}, definitely the restrictions imposed in that paper also apply here, i.e. the near region condition
\be \label{con.near.r}
r\omega \ll 1\,,
\ee 
and case where the wavelength of scalar perturbations is larger than the radius of curvature
\be \label{con.low.f}
M\omega \ll 1\,.
\ee 
The last condition is sometime also called as the low frequency limit. In addition to these conditions, we also need to impose further restriction in regard to the black hole's acceleration, which will allow us to reveal the hidden conformal symmetry in the next section. In this paper, we consider a pair of black holes which are slowly accelerating away from each other, i.e.
\be \label{con.slow.a}
\alpha^2 r^2 \ll 1\,.
\ee 
Furthermore, the near region and low frequency limits yield $\alpha^2 r^2$ is well approached by $\alpha^2 r_+^2$, i.e. it is relatively constant. 

Consequently, it yields
\[
Q \approx Q_ +   \equiv k_+\left( {r^2  - 2Mr + a^2 } \right)
\]
where $k_+ = {1 - \alpha ^2 r_ + ^2 }$. This approximation simplifies the discussion, since instead of dealing with five poles in the original radial equation (\ref{eq.radial}), now we have only three left\footnote{As in the normal scalar probe radial equation in Kerr background, the singularities are at $r=r\pm$ and $r=\infty$.} as in the non-accelerating Kerr black holes case \cite{Castro:2010fd}. To the best of our knowledge, there is no work presented in literature discussing hidden conformal symmetry in the style of \cite{Castro:2010fd}, where the near region and low frequency radial equation contains more than three poles. For example, in revealing the hidden conformal symmetry of Kerr-AdS spacetime \cite{Chen:2011wm}, where the function which gives the horizon locations of the black hole is quartic instead of quadratic like in the generic Kerr case \cite{Castro:2010fd}, the authors also prefer to approach the quartic function to be a quadratic one.

Using the above approaches, now the radial equation (\ref{eq.radial}) can be expressed as
\be 
\left( {\partial _r \Delta \partial _r - \frac{2r\alpha ^2 \left( {r - M} \right)}{k_+} + \frac{{ \left( {\left( {r^2  + a^2 } \right)\omega'  - am'\Xi} \right)^2 }}{\Delta } -  \lambda } \right)R\left( r \right) = 0\,.
\ee 
where $\omega '$, $m'$, and $\lambda$ are defined as $k_+^{-1}\omega$, $k_+^{-1}m$, and $k_+^{-1} {\tilde \lambda}$ respectively. Furthermore, in the near region, low frequency for scalar, and slow acceleration conditions allow one to approximately rewrite the last equation as
\be \label{eq.radial.app}
\left[ {\partial _r \Delta \partial _r  + \frac{{\left( {2Mr_ +  \omega ' - am'\Xi} \right)^2 }}{{\left( {r - r_ +  } \right)\left( {r_ +   - r_ -  } \right)}} - \frac{{\left( {2Mr_ -  \omega ' - am'\Xi} \right)^2 }}{{\left( {r - r_ -  } \right)\left( {r_ +   - r_ -  } \right)}}} \right]R\left( r \right) = l\left( {l + 1} \right)R\left( r \right)\,,
\ee
where we have assigned $\lambda = l\left(l+1\right)$. This is just the radial equation whose Laplacian in the right hand side can be written as the $SL(2,R)_L\times SL(2,R)_R$ squared Casimir operators in \cite{Guica:2008mu}. So, the hidden conformal symmetry certainly exists in the case of the slowly accelerating Kerr black holes. Nevertheless, new constants related to the spacetime properties which appear in the Laplacian, i.e. $k_+$ and $\Xi$, affect the resulting left and right movers temperatures after matching this Laplacian and the constructed squared Casimir. In establishing the Kerr/CFT holography, these temperatures play an important role in recovering the Bekenstein-Hawking entropy by using the 2D CFT Cardy formula.

\section{Hidden conformal symmetry in the wave equation}\label{sec.Hidden}

\subsection{Generic case}

The conformal symmetry hidden in an equation like (\ref{eq.radial.app}) has been widely discussed in the literature \cite{Compere:2012jk,Bredberg:2011hp,Chen:2010as,Chen:2010fr,Chen:2010ywa,Chen:2011kt,Wang:2010qv}. It is argued, that the conformal symmetry of a rotating black hole in non-extremal case is not in the symmetry of spacetime geometry near the horizon, but can be found in the corresponding equation of motion for a scalar probe. This opens the possibility that a 2D CFT could be still the dual description for a non-extremal Kerr black holes. In the original work by Castro et. al. \cite{Castro:2010fd}, the authors constructed the hidden conformal symmetry for a generic Kerr black hole, which laid the foundation for a non-chiral 2D CFT as the dual description for such black holes. In this section, we just highlight the method by Castro et. al. in establishing the conformal symmetry, with some small subtleties in regards to the slow acceleration correction and angular coordinate scale factor in removing the nodal singularity. The obtained left and right movers temperatures, after matching the Laplacian and squared Casimir, are exactly those which are needed in the Cardy formula with the central charge is that of \cite{Astorino:2016xiy} or the one obtained in appendix \ref{sec.app1} using an alternative calculation.

The starting point is defining the conformal coordinates in terms of the Hong-Teo type ones, $\left\{ {t,r,\phi } \right\}$, which are involved in radial equation (\ref{eq.radial.app}), i.e.
\[
\omega ^ +   = \sqrt {\frac{{r - r_ +  }}{{r - r_ -  }}} \exp \left( {2\pi T_R \phi  + 2\pi n_R t} \right)\,,
\]
\be 
\omega ^ -   = \sqrt {\frac{{r - r_ +  }}{{r - r_ -  }}} \exp \left( {2\pi T_L \phi  + 2\pi n_L t} \right)\label{coord.conf}\,,
\ee 
\[
y = \sqrt {\frac{{r_ +   - r_ -  }}{{r - r_ -  }}} \exp \left( {\pi \left( {T_R  + T_L } \right)\phi  + \pi \left( {n_R  + n_L } \right)t} \right)\,.
\]
Using these conformal coordinates, we then define the following generators
\be 
H_ +   = i\frac{\partial }{{\partial \omega ^ +  }}~,~
H_0  = i\left( {\omega ^ +  \frac{\partial }{{\partial \omega ^ +  }} + \frac{y}{2}\frac{\partial }{{\partial y}}} \right)\label{genH}~,~
H_ -   = i\left( {\left( {\omega ^ +  } \right)^2 \frac{\partial }{{\partial \omega ^ +  }} + \omega ^ +  y\frac{\partial }{{\partial y}} - y^2 \frac{\partial }{{\partial \omega ^ -  }}} \right) \,,
\ee
and
\be 
\bar H_ +   = i\frac{\partial }{{\partial \omega ^ -  }}~,~ 
\bar H_0  = i\left( {\omega ^ -  \frac{\partial }{{\partial \omega ^ -  }} + \frac{y}{2}\frac{\partial }{{\partial y}}} \right) \label{genHbar}~,~
\bar H_ -   = i\left( {\left( {\omega ^ -  } \right)^2 \frac{\partial }{{\partial \omega ^ -  }} + \omega ^ -  y\frac{\partial }{{\partial y}} - y^2 \frac{\partial }{{\partial \omega ^ +  }}} \right)\,.
\ee
The two set of generators above satisfy the $SL(2,R)$ algebra, $\left[ {H_ \pm  ,H_0 } \right] =  \pm iH_ \pm  $, and $\left[ {H_ +  ,H_ -  } \right] = 2iH_0 $. The same commutations also apply to the bar version of generators, $\left\{ {\bar H_ +  ,\bar H_0 ,\bar H_ -  } \right\}$. 

Using the vectors in (\ref{genH}) and (\ref{genHbar}), we can construct the squared Casimirs
\be\label{Casimir2}
{\cal H}^2 = - H_0^2  + \frac{1}{2}\left( {H_ +  H_ -   + H_ -  H_ +  } \right) \,,
\ee
\be\label{Casimir2bar}
{\bar {\cal H}}^2 =  - \bar H_0^2  + \frac{1}{2}\left( {\bar H_ +  \bar H_ -   + \bar H_ -  \bar H_ +  } \right)\,,
\ee
which commute to all corresponding generators, namely
\be 
\left[ {{\cal H}^2,H_k } \right] = \left[ {{\bar {\cal H}}^2,\bar H_k } \right] = 0
\ee
for $k = +,-,0$. Based on these commutations, one can infer that these squared Casimir is related to some conserved quantities in the system, which turns out to be $\lambda = l\left(l+1\right)$.

To proceed further, one can write down the explicit form of squared Casimir \cite{Chen:2011kt}
\[
{\cal H}^2 = {\bar {\cal H}}^2 = \partial _r \left( {\left( {r - r_ +  } \right)\left( {r - r_ +  } \right)} \right) \partial _r 
\]
\[
- \frac{{\left( {r_ +   - r_ -  } \right)}}{{\left( {r - r_ +  } \right)}}\left( {\frac{{\pi \left( {T_L  + T_R } \right)\partial _t  - \left( {n_L  + n_R } \right)\partial _\phi  }}{{4\pi \left( {T_L n_R  - T_R n_L } \right)}}} \right)^2  + \frac{{\left( {r_ +   - r_ -  } \right)}}{{\left( {r - r_ -  } \right)}}\left( {\frac{{\pi \left( {T_L  - T_R } \right)\partial _t  - \left( {n_L  - n_R } \right)\partial _\phi  }}{{4\pi \left( {T_L n_R  - T_R n_L } \right)}}} \right)^2 \,,\]
which after acting on the angular independent part of the wave function $S\left(\theta\right)^{-1} \Phi = e^{-i\left(\omega t -m\phi\right)} R\left(r\right)$, it gives

\[
{\cal H}^2 R\left( r \right) = {\bar {\cal H}}^2 R\left( r \right) = \partial _r \left( {\left( {r - r_ +  } \right)\left( {r - r_ +  } \right)\partial _r R\left( r \right)} \right)
\]
\be \label{sqrd.Casimir.res}
+ \frac{{\left( {r_ +   - r_ -  } \right)}}{{\left( {r - r_ +  } \right)}}\left( {\frac{{\pi \left( {T_L  + T_R } \right)\omega  + \left( {n_L  + n_R } \right)m}}{{4\pi \left( {T_L n_R  - T_R n_L } \right)}}} \right)^2  - \frac{{\left( {r_ +   - r_ -  } \right)}}{{\left( {r - r_ -  } \right)}}\left( {\frac{{\pi \left( {T_L  - T_R } \right)\omega  + \left( {n_L  - n_R } \right)m}}{{4\pi \left( {T_L n_R  - T_R n_L } \right)}}} \right)^2 R\left( r \right)\,.
\ee 
Matching the this squared Casimir (\ref{sqrd.Casimir.res}) to the Laplacian in (\ref{eq.radial.app}) gives the results
\be \label{TlTr.nonext}
T_L  = \frac{{\left( {r_ +   + r_ -  } \right)k_ +  }}{{4\pi a\Xi}}\, , \,T_R  = \frac{{\left( {r_ +   - r_ -  } \right)k_ +  }}{{4\pi a\Xi}}\, , \,n_L  =  - \frac{{k_ +  }}{{4M}} \,.
\ee 
Definitely, these parameters $T_{R,L}$ and $n_L$ reduce to that of generic Kerr \cite{Castro:2010fd} after the non-accelerating consideration $\alpha =0$ is taken, which yields $\Xi = 1$ and $k_+ = 1$.

\subsection{Extremal case}

Subsequently, short after the proposal of hidden conformal symmetry for non-extremal Kerr black holes, its extremal version was presented in \cite{Chen:2010fr}. The authors of \cite{Chen:2010fr} managed to reveal the $SL\left(2,R\right)$ symmetry in the equation of motion for probes in the background of extremal Kerr black holes. Note that the extremal version of hidden conformal symmetry cannot be taken simply by setting $r_+ - r_-\to 0$ from the generators in (\ref{genH}) and (\ref{genHbar}). Interestingly, the extremal Kerr/CFT correspondence based on the hidden conformal symmetry \cite{Chen:2010fr} resembles the type of holographic dual theory which is discussed in \cite{Guica:2008mu}, in the sense that it is chiral instead of non-chiral as one in the non-extremal case\footnote{Based on the non-zero parameters $T_L$ and $n_L$, instead of $T_{L,R}$ and $n_{L,R}$ of a non-chiral one.}.

In this section, following the work in \cite{Chen:2010fr}, we show that a slowly accelerating extremal Kerr black holes also possess hidden conformal symmetry, just like its non-extremal counterpart. As the black hole is in the extreme condition, i.e. $r_+ = r_-$, the near region and low frequency scalar (\ref{eq.radial.app}) reads
\be \label{eq.extremal}
\left( {\partial _r \left( {r - r_ +  } \right)^2 \partial _r  + \frac{{4M^2\omega '\left( {2M \omega ' - m'\Xi} \right)}}{{r - r_ +  }} + \frac{{M^2\left( {2M \omega'  - m'\Xi} \right)^2 }}{{\left( {r - r_ +  } \right)^2 }}} \right)R\left( r \right) = \lambda R\left( r \right)\,.
\ee 
The mapping from conformal coordinates $\left\{ {\omega ^ +  ,\omega ^ -  ,y} \right\}$ to the Hong-Teo type ones $\left\{ {t,r,\phi } \right\}$ in extremal case is quite different compared to the transformation in non-extremal case (\ref{coord.conf}). It reads \cite{Chen:2010fr}
\[
\omega ^ +   = \frac{c}{2}\left( {\frac{\phi }{a} - \frac{1}{{r - r_ +  }}} \right)\,, \]\be \label{coord.conf.ext}
\omega ^ -   = \frac{1}{2}\exp \left( {2\pi T_L \phi  + 2n_L t} \right) - \frac{1}{c} \,,\ee
\[ 
y = \sqrt {\frac{c}{{2\left( {r - r_ +  } \right)}}} \exp \left( {\pi T_L \phi  + n_L t} \right)\,.
\]
Using the mapping (\ref{coord.conf.ext}), one can also construct the vectors (\ref{genH}) and (\ref{genHbar}). Then, using these vectors, the squared Casimir operators (\ref{Casimir2}) and (\ref{Casimir2bar}) can be built in showing the hidden conformal symmetry for the extremal and slowly accelerating Kerr black hole. The reading of these squared Casimirs in Hong-Teo type coordinate is
\[
{\cal H}^2  = {\bar {\cal H}}^2  = \partial _r \left( {r - r_ +  } \right)^2 \partial _r  - \frac{{a^2 \left( {\pi T_L \partial _t  - n_L \partial _\phi  } \right)^2 }}{{n_L^2 \left( {r - r_ + } \right)^2 }} - \frac{{a\left( {\pi T_L \partial _t^2  - n_L \partial _\phi  \partial _t } \right)}}{{n_L^2 \left( {r - r_ + } \right)}}\,.
\]
Acting this operator to the angular independent of the wave function $e^{ - i\left( {\omega t - m\phi } \right)} R\left( r \right)$ results
\be \label{Cas.op.ext}
\partial _r \left( {\left( {r - r_ +  } \right)^2 \partial _r R\left( r \right)} \right) + \left( {\frac{{a^2 \left( {\pi T_L \omega  + n_L m} \right)^2 }}{{n_L^2 \left( {r - r_ + } \right)^2 }} + \frac{{a\left( {\pi T_L \omega ^2  + 6n_L m\omega } \right)}}{{n_L^2 \left( {r - r_ + } \right)}}} \right)R\left( r \right)\,,
\ee
whose eigen value is the separation constant $\lambda$ in the r.h.s. of eq. (\ref{eq.extremal}). Finally, using the matching trick between the operator (\ref{Cas.op.ext}) and Laplacian in l.h.s. of eq (\ref{eq.extremal}), the following constant quantities can be identified as
\be \label{Tl.extrem}
T_L  = \frac{{k_ +  }}{{2\pi \Xi }}~,~n_L  =  - \frac{{k_ +  }}{{4M}}\,.
\ee 
As it is expected, when the non-accelerating case is considered, i.e. $\alpha=0$, thence the parameters in (\ref{Tl.extrem}) reduce to those in extremal Kerr \cite{Guica:2008mu,Chen:2010fr}. Interestingly, the left mover temperature presented in (\ref{Tl.extrem}) is just the Frolov-Thorne\footnote{In \cite{Guica:2008mu,Astorino:2016xiy}, this temperature is obtained by constructing the Frolov-Thorne vacuum and then take the limit $T_H\to 0$ of the resulting Boltzmann factor.} temperature which is used in computing the dual entropy for an extremal Kerr black hole \cite{Guica:2008mu}.

\section{Holography}\label{sec.holography}

In this section we present some evidences for the duality between a slowly accelerating Kerr black hole and a 2D CFT. They are the dual calculations of Bekenstein-Hawking entropy and absorption cross section. The discussions here cover both the extremal and non-extremal cases.

\subsection{Dual entropy}

Now let us compute the microscopic entropy for a slowly accelerating Kerr black hole. In the original Kerr/CFT correspondence proposal \cite{Guica:2008mu}, the central charge is computed by restricting the analysis into an extremal case only. The Bekenstein-Hawking entropy then can be recovered using the chiral Cardy formula, where the associated temperature is the non-zero Frolov-Thorne temperature near the horizon. It is still an unclear business to get the central charge for the non-extremal Kerr black hole using the ASG method, even though some attempts can be found in the literature using another approaches \cite{Carlip:2011ax,ChangYoung:2012kd,Majhi:2014lka}. This problem can be understood from the fact that the near horizon of non-extremal black holes takes the Rindler form instead of a warped or squashed AdS, hence ASG method cannot be applied. 

Nevertheless, one can assume that a non-extremal central charge which smoothly reduce to that in the extremal case does exist. Assuming this smoothness to the case of accelerating Kerr geometry, one can expect the central charge in the non-extremal condition reads
\be\label{central.charge.nonex}
c = \frac{{12J}}{{k_ + ^2 }}\,,
\ee
which reduces to (\ref{central.charge.final}) after setting $r_+ = r_-$. Moreover, one can also argue that the left and right movers theories are characterized by the same central charge (\ref{central.charge.nonex}). Thence, using the Cardy formula from a product of the left and right mover theories, 
\be 
S_{Cardy}  = \frac{{\pi ^2 }}{3}\left( {c_L T_L  + c_R T_R } \right)\,,
\ee
and left and right temperatures are given in (\ref{TlTr.nonext}), one can recover the Bekenstein-Hawking entropy for accelerating black holes
\be 
S_{Cardy}  = S_{BH}  = \frac{{A_{BH} }}{4} = \frac{2\pi M r_+}{1-\alpha^2 r_+^2}\,.
\ee
As the black holes are in extremal state, the chiral Cardy formula
\be 
S_{Cardy}  = \frac{{\pi ^2 }}{3}c_L T_L 
\ee
where the central charge $c_L$ is the extremal one (\ref{central.charge.final}) and the left temperature (\ref{Tl.extrem}), again recovers the extremal Bekenstein-Hawking entropy for an accelerating black hole
\be 
S_{BH}  = \frac{{2\pi M^2 }}{{1 - \alpha ^2 M^2 }}\,.
\ee

\subsection{Scattering process and 2D CFT}\label{sec.dualScattering}

Another evidence that Kerr black holes are dual to a 2D CFT is the agreement, up to some factors, between the absorption cross section formula obtained from gravitational and 2D CFT formulas. In this section, we will explore this aspect for both non-extremal and extremal cases. In the non-extremal case, we can start by writing down the equation (\ref{eq.radial.app}) in term of a new dimensionless variable 
\be 
z = \frac{{r - r_ +  }}{{r - r_ -  }}\,.
\ee 
The resulting equation is
\be 
\left( {\left( {1 - z} \right)\partial _z z\partial _z  + \frac{{K_ + ^2 }}{z} - K_ - ^2  - \frac{{l\left( {l + 1} \right)}}{{1 - z}}} \right)R\left( z \right) = 0\,,
\ee 
where the constants $K_ \pm$ are given by
\be
K_ \pm   = \frac{{2Mr_ \pm  \omega ' - am'\Xi }}{{r_ +   - r_ -  }}\,.
\ee 
Solutions to the last equation can be written as the superposition of the ingoing and outgoing modes
\be  
R_{out}  = z^{iK_ +  } \left( {z - 1} \right)^{-l}~  _2 F_1 \left( {{-l}  + i\left( {K_ +   - K_ -  } \right),{-l}  + i\left( {K_ +   + K_ -  } \right);1 + 2iK_ +  ;z} \right)\,,
\ee
and
\be 
R_{in}  = z^{ - iK_ +  } \left( {z - 1} \right)^{-l}~  _2 F_1 \left( {{-l}  - i\left( {K_ +   - K_ -  } \right),{-l}  - i\left( {K_ +   + K_ -  } \right);1 - 2iK_ +  ;z} \right)\,.
\ee
Considering the outer part of near region, i.e. $r\gg M$ but still subjects to $r\omega \ll 1$, allows one to have
\be \label{Rin.far}
R_{in}  \sim Ar^{l}  + Br^{ - l -1} \,,
\ee 
where
\[
A = \frac{{\Gamma \left( { - 2l - 1} \right)\Gamma \left( {1 - 2iK_ +  } \right)}}{{\Gamma \left( {1 + l - i\left( {K_ +   - K_ -  } \right)} \right)\Gamma \left( {1 + l - i\left( {K_ +   + K_ -  } \right)} \right)}}\,,
\]
and
\[
B = \frac{{\Gamma \left( {2l + 1} \right)\Gamma \left( {1 - 2iK_ +  } \right)}}{{\Gamma \left( { - l - i\left( {K_ +   - K_ -  } \right)} \right)\Gamma \left( { - l - i\left( {K_ +   + K_ -  } \right)} \right)}}\,.
\]

To show the possibility that a 2D CFT might be the dual description for a Kerr black hole, from the absorption cross section point of view, we need to identify some parameters which appear in the expression of (\ref{Rin.far}). To proceed, we need to recall the first law of thermodynamics for an accelerating and rotating black hole, namely
\be\label{first.law.thermo}
T_H \delta S_{BH}  = \delta M - \Omega _H \delta J\,.
\ee
Note that, in writing down the last relation we limit our discussion of accelerating black hole whose acceleration comes from a semi infinite cosmic string with a fixed tension\footnote{A recent work \cite{Appels:2017xoe} incorporates the change of string tension into the thermodynamical relation of an accelerating black hole.}, hence the change of black hole entropy depends solely on the variation of black hole mass and angular momentum.

Assuming the holographic relation between the physics of black hole and a 2D CFT with $SL\left( {2,R} \right)_L  \times SL\left( {2,R} \right)_R$ symmetry, then the relation $\delta S_{BH}  = \delta S_{CFT} $ should hold. Consequently, this yields
\be \label{dSBHdSCFT}
\frac{{\delta M - \Omega _H \delta J}}{{T_H }} = \frac{{\delta E_L }}{{\delta T_L }} + \frac{{\delta E_R }}{{\delta T_R }}\,.
\ee
Now let us identify $ \delta M = \omega ' $, $ \delta J = m' $, $ \delta E_L  = \omega _L  $, and $ \delta E_R  = \omega _R  $, which then lead us the expressions for the left and right movers frequencies as
\be 
\omega _L  = \frac{{2M^2 \omega }}{{a\Xi }}~,~\omega_R = \omega_L - m\,.
\ee
Furthermore, let us assign the the left and right conformal weights $h_L  = h_R  = h = l+1$.

Using the solution (\ref{Rin.far}), the retarded Green function can be written as
\be 
G_R  \sim \frac{B}{A }\propto \sin \left( {\pi h_L  + i\frac{{\omega _L }}{{2T_L }}} \right)\left| {\Gamma \left( {h_L  + i\frac{{\omega _L }}{{2\pi T_L }}} \right)} \right|^2 \sin \left( {\pi h_R  + i\frac{{\omega _R }}{{2T_R }}} \right)\left| {\Gamma \left( {h_R  + i\frac{{\omega _R }}{{2\pi T_R }}} \right)} \right|^2 \,.
\ee 
This expression is in agreement with the Euclidean correlator in 2D CFT, namely
\[
G_E  \sim T_L^{2h_L  - 1} \exp \left[ {i\frac{{{\tilde \omega} _L }}{{2T_L }}} \right]\Gamma \left( {h_L  + \frac{{{\tilde \omega} _L }}{{2\pi T_L }}} \right)\Gamma \left( {h_L  - \frac{{{\tilde \omega} _L }}{{2\pi T_L }}} \right)
\]
\be 
~~~~~~~~~~~~~~~ \times T_R^{2h_R  - 1} \exp \left[ {i\frac{{\omega _R }}{{2T_R }}} \right]\Gamma \left( {h_R  + \frac{{\omega _R }}{{2\pi T_R }}} \right)\Gamma \left( {h_R  - \frac{{\omega _R }}{{2\pi T_R }}} \right)\,,
\ee
where $\tilde \omega$ is the Euclidean frequencies, i.e. ${\tilde \omega} = i\omega$. Furthermore, from this retarded Green function one can obtain the absorption cross section trough the relation
\be 
\sigma _{abs}  \sim {\mathop{\rm Im}\nolimits} G_R  \propto \sinh \left( {\frac{{\omega _L }}{{2T_L }} + \frac{{\omega _R }}{{2T_R }}} \right)\left| {\Gamma \left( {h_L  + i\frac{{\omega _L }}{{2\pi T_L }}} \right)} \right|^2 \left| {\Gamma \left( {h_R  + i\frac{{\omega _R }}{{2\pi T_R }}} \right)} \right|^2 \,,
\ee 
which has a dual 2D CFT expression
\be 
\sigma _{CFT}  \sim T_L^{2h_L  - 1} T_R^{2h_R  - 1} \sinh \left( {\frac{{\omega _L }}{{2T_L }} + \frac{{\omega _R }}{{2T_R }}} \right)\left| {\Gamma \left( {h_L  + i\frac{{\omega _L }}{{2\pi T_L }}} \right)} \right|^2 \left| {\Gamma \left( {h_R  + i\frac{{\omega _R }}{{2\pi T_R }}} \right)} \right|^2 \,.
\ee 
Recall that the last formula is the finite temperature absorption cross section in a 2D CFT. At this point, we have established the non-extremal Kerr/CFT correspondence for an accelerating Kerr black hole by studying the scattering process. This adds to the entropy matching in the previous section as some evidences for the holography between accelerating Kerr black holes and a 2D CFT.

Now let us turn the discussion to the extremal black holes. It turns out that the coordinate transformation
\be \label{ztorExt}
z = \frac{{2iM\left( {2M\omega ' - m'\Xi } \right)}}{{r - M}}
\ee 
leads to the reading of (\ref{eq.extremal}) as the following
\be \label{eq.Kummer.ext}
\frac{{d^2 R\left( z \right)}}{{dz^2 }} + \left( { - \frac{1}{4} + \frac{p}{z} + \frac{{1 - 4q^2 }}{{4z^2 }}} \right)R\left( z \right) = 0\,,
\ee
where $p =  - 2iM\omega k_+^{-1}$ and $q^2 = \tfrac{1}{4} + \lambda $. The differential equation (\ref{eq.Kummer.ext}) can be solved by
\be 
R\left( z \right) = c_ +  e^{ - \frac{z}{2}} z^{\frac{1}{2} + q} K\left( {\frac{1}{2} + q - p,1 + 2q;z} \right) + c_ -  e^{ - \frac{z}{2}} z^{\frac{1}{2} - q} K\left( {\frac{1}{2} - q - p,1 - 2q;z} \right)\,,
\ee
where $c_\pm$ are some constants to be fixed by the boundary condition (\ref{bc.cpcm}), and $K\left( {\mu ,\nu ;z} \right)$ is the Kummer function\footnote{This function solves $z\frac{{d^2 K}}{{dz^2 }} + \left( {\nu  - z} \right)\frac{{dK}}{{dz}} = \mu K$.}. Near the black hole horizon $r_+$, i.e. $z\to\infty$, the Kummer function asymptotically can be approached by
\be 
K\left( {\mu ,\nu ,z} \right) \sim \frac{{\Gamma \left( \nu  \right)e^{ - i\mu \pi } z^{ - \mu } }}{{\Gamma \left( {\nu  - \mu } \right)}} + \frac{{\Gamma \left( \nu  \right)e^z z^{\mu  - \nu } }}{{\Gamma \left( \mu  \right)}}\,.
\ee
Moreover, requiring the wave must be purely ingoing near horizon constraints the constants $c_{\pm}$ to be related as
\be \label{bc.cpcm}
\frac{{\Gamma \left( {\tfrac{1}{2} - q - p} \right)c_ +  }}{{\Gamma \left( {1 - 2q} \right)}} + \frac{{\Gamma \left( {\tfrac{1}{2} + q - p} \right)c_ -  }}{{\Gamma \left( {1 + 2q} \right)}} = 0\,.
\ee

In the asymptotic $r\to \infty$, $z\to 0$, which consequently implies $K\left(\mu , \nu , z\right)\to 1$, we have
\be 
R\left( r \right) \sim c_ +  r^{ - h_L }  + c_ -  r^{ - h_L  - 1} \,,
\ee
where the left mover conformal weight for scalar operator reads
\be 
h_L  = \frac{1}{2} + \sqrt {\lambda  + \frac{1}{4}} \,.
\ee
Furthermore, if one considers to write 
\be 
p = i\frac{{\omega _L }}{{2\pi T_L }}\,,
\ee
where $\omega _L  = 4\pi M\omega T_L k_ + ^{ - 1}$, thence the retarded Green function obtained from the ratio of constants $c_\pm$ reads
\be \label{Green.ext}
G_R  \sim \frac{{c_ +  }}{{c_ -  }} \propto \frac{{\Gamma \left( {h_L  - i\frac{{\omega _L }}{{2\pi T_L }}} \right)}}{{\Gamma \left( {1 - h_L  - i\frac{{\omega _L }}{{2\pi T_L }}} \right)}} \propto \sin \left( {\pi \left( {h_L  + i\frac{{\omega _L }}{{2\pi T_L }}} \right)} \right)\Gamma \left( {h_L  - i\frac{{\omega _L }}{{2\pi T_L }}} \right)\Gamma \left( {h_L  + i\frac{{\omega _L }}{{2\pi T_L }}} \right)\,.
\ee
The expression of this Green function (\ref{Green.ext}) in terms of Euclidean frequency $\tilde \omega$ agrees to the 2D CFT correlator
\be 
G_E  \sim T_L^{2h_L  - 1} \exp \left[ {i\frac{{\tilde \omega _L }}{{2T_L }}} \right]\Gamma \left( {h_L  - \frac{{\tilde \omega _L }}{{2\pi T_L }}} \right)\Gamma \left( {h_L  + \frac{{\tilde \omega _L }}{{2\pi T_L }}} \right)\,.
\ee
Furthermore, the absorption cross section from this retarded Green function can be written as
\be 
\sigma_{grav}  \sim {\mathop{\rm Im}\nolimits} G_R  \propto \sinh \left( {\frac{{\tilde \omega _L }}{{2T_L }}} \right)\left| {\Gamma \left( {h_L  - i\frac{{\tilde \omega _L }}{{2\pi T_L }}} \right)} \right|^2 \,,
\ee
whereas the 2D CFT counterpart, namely the finite temperature absorption cross section reads
\be 
\sigma _{CFT}  \sim T_L^{2h_L  - 1} \sinh \left( {\frac{{\tilde \omega _L }}{{2T_L }}} \right)\left| {\Gamma \left( {h_L  - i\frac{{\tilde \omega _L }}{{2\pi T_L }}} \right)} \right|^2 \,.
\ee
Now we can observe that the last two expressions match each other up to some factors, which supports the holographic relation between an extremal accelerating Kerr black hole and 2D CFT.

The results presented in here and the last section may suggest that the taking the slowly accelerating limit of Kerr spacetime is just to make the spacetime under consideration to coincide with the generic Kerr. However, from the Kretschmann scalar for an accelerarting Kerr spacetime calculated in app. \ref{sec.app3}, one can observe that the limit $\alpha^2 r^2 \ll 1$ does not lead to the Kretschmann scalar for generic Kerr expression. This fact tells us that the generic Kerr spacetime cannot be mapped by using any coordinate or gauge transformation to obtain the slowly accelerating limit of (\ref{metric.acc.Kerr.ori}).

\section{Conclusion and discussion}\label{sec.CON}

To conclude, we have generalized the works presented in \cite{Castro:2010fd,Chen:2010fr} to the case of a pair rotating black holes which are slowly accelerating away each other. The most important part in our study is the existence of hidden conformal symmetry in the slowly accelerating Kerr spacetime, as presented in section \ref{sec.Hidden}. Using the left and right temperatures appearing in coordinate transformations (\ref{coord.conf}) and (\ref{coord.conf.ext}), and an assumption that the associated central charge is smoothly obtained from the extremal case, in section \ref{sec.holography} we show that the Bekenstein-Hawking entropy of the slowly accelerating Kerr black holes can be recovered using Cardy formula in 2D CFT. This is in line with the proposal of Kerr/CFT holography \cite{Guica:2008mu}. Furthermore, the matching of absorption cross section from the gravitational calculation and the 2D CFT analysis supports the conjecture that there exists some relations between the slowly accelerating Kerr black holes to a 2D CFT. Indeed, similar conjectures have been proposed for many rotating black holes \cite{Compere:2012jk}, but not the accelerating case. 

Since its original proposal, the hidden conformal symmetry of Kerr black hole \cite{Castro:2010fd} has been extended to many more complicated system discussing rotating black holes in various theories \cite{Compere:2012jk}. For example in the case of Einstein-Maxwell \cite{Chen:2010ywa} and low energy heterotic string \cite{Ghezelbash:2013uga}, the presence of electromagnetic field enriches the hidden conformal symmetry of the system, i.e. the existence of rotational $J$ and charge $Q$ pictures. We conjecture that this should be also the case for a slowly accelerating Kerr-Newman-(A)dS black hole, provided that the minimally coupled Klein-Gordon equation in this background is still separable\footnote{To the best of our knowledge, this investigation is still absent in literature.}. Proving the separability of minimally coupled Klein-Gordon equation in accelerating Kerr-Newman background, and also checking this conjecture would be some challenging future works.

\section*{Acknowledgement}
 
I thank Zoltan Bajnok for his supports and encouragement, Michael C. Abbott and Bobby Gunara for the discussions. I also thank my colleagues from Physics Department of Parahyangan Catholic University, and anonymous referees for his/her comments. This work is supported by Tempus Public Foundation under Grant No. BE AK2017/162458.

\appendix
\section{Central charge from stretched horizon method}\label{sec.app1}

In \cite{Carlip:2011ax}, Carlip showed how to get the central charge associated to the conformal symmetries of the near horizon Kerr black holes without taking the near horizon of an extremal Kerr (NHEK) geometry first, as in \cite{Guica:2008mu}. Furthermore, the Carlip's approach, which is sometime referred as the stretched horizon method, does not require one to impose a boundary condition for the metric perturbation $h_{\mu \nu }  = {\cal L}_\zeta  g_{\mu \nu } $. Here, $\zeta^{\mu}$ is the asymptotic symmetric group (ASG) diffeomorphism generator. Nevertheless, the authors of \cite{Chen:2011wm} showed that both techniques are equivalent, and surely one will end with the same results by using either ways. The author of \cite{Astorino:2016xiy} managed to obtain the central charge related to the accelerating Kerr black holes by using the ASG method \cite{Guica:2008mu}. This appendix is dedicated to highlight the alternative computation regarding to the same problem.

We start by writing the line element (\ref{metric.acc.Kerr.ori}) into the ADM form which reads
\be \label{metric.acc.Kerr.ADM}
ds^2  =  - N^2 dt^2  + h_{rr} dr^2  + h_{\theta \theta } d\theta ^2  + h_{\phi \phi } \left( {d\phi  + N^\phi  dt} \right)^2 \,,
\ee
where
\[
N^2  = \frac{{\rho ^2 PQ}}{{\Omega ^2 \left( {P\left( {r^2  + a^2 } \right)^2  - Qa^2 \sin ^2 \theta } \right)}}\,,
\]
\[
h_{rr}  = \frac{{\rho ^2 }}{{\Omega ^2 Q}}\,,
\]
\[
h_{\theta \theta }  = \frac{{\rho ^2 }}{{\Omega ^2 P}}\,,
\]
\[
h_{\phi \phi }  = \frac{{\sin ^2 \theta \left( {P\left( {r^2  + a^2 } \right)^2  - Qa^2 \sin ^2 \theta } \right)}}{{C^2\Omega ^2 \rho ^2 }}\,,
\]
and
\[
N^\phi   = \frac{{aC\left( {Q - P\left( {r^2  + a^2 } \right)} \right)}}{{P\left( {r^2  + a^2 } \right)^2  - Qa^2 \sin ^2 \theta }}\,.
\]
The area and angular velocity at the horizon for this accelerating black hole in extremal condition are given by
\be 
A_{BH}  = \frac{{8\pi a^2 }}{C\left({1 - \alpha ^2 a^2 }\right)}\,,
\ee
and
\be 
\Omega _H  = \frac{C}{{2a}}\,,
\ee
respectively. Then, following \cite{Chen:2011wm,Carlip:2011ax}, the central charge can be obtained using te formula
\be \label{central.charge.stretched}
c = \frac{{3\gamma \delta A_{BH} }}{{2\pi \beta }}
\ee
where
\[
\gamma  = \left. {\sqrt {h_{rr} } \left( {r - r_ +  } \right)} \right|_{r - r_ +  } \,,
\]
\[
\delta  = \left. {\frac{{N^\phi   + \Omega _H }}{{r - r_ +  }}} \right|_{r - r_ +  } \,,
\]
and
\[
\beta  = \left. {\frac{N}{{r - r_ +  }}} \right|_{r - r_ +  } \,.
\]
Note that all these $\beta$, $\gamma$, and $\delta$ functions are evaluated in extremal case, i.e. $M=a$. Moreover, the $Q$ function in the equations above which resembles $\Delta = \left(r-r_+\right)^2$ for Kerr case vanishes at $r=r_+$. Evaluating the formula (\ref{central.charge.stretched}) for the line element (\ref{metric.acc.Kerr.ADM}) in extremal case $M=a$ and at the horizon $r=a$ gives
\be\label{central.charge.final} 
c = \frac{{12a^2 }}{{\left( {1 - \alpha ^2 a^2 } \right)^2 }}\,.
\ee
This is the central charge associated to the conformal symmetry in the near horizon of an extremal accelerating Kerr black hole, which is calculated using ASG method in \cite{Astorino:2016xiy}.

\section{Hawking temperature of an accelerating Kerr black hole from tunneling method}\label{sec.app2}

As in the case of non-accelerating black hole, an accelerating black hole not only absorbs objects around it, but also emits particle in the form of Hawking radiation. Normally, the Hawking temperature for this black hole is obtained by  calculating its surface gravity first. As an alternative way to this method, one can also employ the tunneling picture \cite{Parikh:1999mf} which sometime is considered to be more intuitive. This tunneling method has been discussed to many cases of black holes, stationary or dynamical cases \cite{Vanzo:2011wq}. As some examples, for Kerr black holes in \cite{Angheben:2005rm,Banerjee:2008cf}, and time dependent black holes in \cite{Siahaan:2009qv}.

For a stationary and spherical symmetric, where the line element is diagonal,
\be 
ds^2  =  - f\left( r \right)dt^2  + \frac{{dr^2 }}{{g\left( r \right)}} + r^2 \left( {d\theta ^2  + \sin ^2 \theta d\phi ^2 } \right)\,,
\ee
the corresponding Hawking temperature is \cite{Banerjee:2008cf}
\be \label{TH}
T_H  = \frac{{\sqrt {f'\left( {r_h } \right)g'\left( {r_h } \right)} }}{{4\pi }}\,,
\ee 
where $r_h$ is horizon position and primed notation denotes the differentiation with respect to the radius. In getting (\ref{TH}), we have considered the radial null geodesic, i.e. $d\theta = d\phi =0$. 

For an accelerating Kerr black hole, by also considering the radial and null geodesic in $\theta =0$ axis\footnote{Again to simplify the analysis.}, the function $f\left(r\right)$ and $g\left(r\right)$ which appear in (\ref{TH}) are 
\be \label{fgACCbh}
f\left( r \right) =  - g_{tt} ~~,~~ g\left( r \right) = g_{rr}^{ - 1} \,,
\ee
from the metric component (\ref{metric.acc.Kerr.fixed.nodal}). Plugging (\ref{fgACCbh}) into the formula (\ref{TH}), we can recover the Hawking temperature for an accelerating Kerr black hole \cite{Astorino:2016xiy},
\be \label{THacc}
T_H  = \frac{{\left( {r_ +   - r_ -  } \right)\left( {1 - \alpha ^2 r_ + ^2 } \right)}}{{4\pi \left( {r_ + ^2  + a^2 } \right)}}\,.
\ee 
As it is expected, this temperature vanishes as the black hole reach its extremal state.

\section{Kretschmann scalar for the accelerating Kerr spacetime}\label{sec.app3}

It is well known that scalar does not change under coordinate transformation. In regard to spacetimes, which may be related by some coordinated transformations, scalar quantities associated to the spacetimes can tell whether two spacetimes are the same but look different due to coordinates choices, or they totally differ each other. Examples for scalars that are related to the geometry of the spacetime are Ricci and Kretschmann scalars, i.e. $R_{\mu}^{\mu}$ and $R_{\alpha \beta \mu \nu } R^{\alpha \beta \mu \nu }$ respectively. Unfortunately, since the system discussed in this paper is vacuum Einstein, Ricci scalar cannot be used to distinguished between the available solutions in the system. However, Kretschmann scalar which is the squared of Riemann curvature tensor appears distinguishable for different solutions in the vacuum Einstein system. Equipped with this property, we can make distinction between accelerating Kerr, Kerr, and Schwarzschild solutions.

For the spacetime (\ref{metric.acc.Kerr.ori}), it can be found that the corresponding Kretschmann scalar reads
\be \label{Kretschmann}
K = R_{\alpha \beta \mu \nu } R^{\alpha \beta \mu \nu }  = \frac{{48M^2 \left( {1 - \alpha rx} \right)^6  }}{{\left( {r^2  + a^2 x^2 } \right)^6 }}\sum\limits_{i = 0}^3 {\left( {ax} \right)^{3 - i} f_i r^i } \sum\limits_{j = 0}^3 {\left( {ax} \right)^{3 - j} \tilde f_j r^j } \,,
\ee
where
\be \label{fkret1}
f_3  =  - \frac{{f_1 }}{3} =  - \frac{{\tilde f_2 }}{3} = \tilde f_0  = 1 + a\alpha \,,
\ee 
and
\be \label{fkret2}
f_0  =  - \frac{{f_2 }}{3} =  - \tilde f_3  =  - \frac{{\tilde f_1 }}{3} = 1 - a\alpha \,.
\ee 
Taking $\alpha\to 0$ in this Kretschmann scalar (\ref{Kretschmann}) gives the one associated to the Kerr spacetime, which reads \cite{Henry:1999rm}
\be \label{KretschmannKerr}
K_{Kerr}  = \frac{{48M^2 \left\{ {\left[ {\left( {r^2  + a^2 x^2 } \right)^2  - \left( {4arx} \right)^2 } \right]\left( {r^2  - a^2 x^2 } \right)} \right\}}}{{\left( {r^2  + a^2 x^2 } \right)^6 }}\,.
\ee
Note that the limit $\alpha^2 r^2 \ll 1$ does not bring the expression (\ref{Kretschmann}) reduces to (\ref{KretschmannKerr}), from which we can infer that the slow acceleration limit, $\alpha^2 r^2 \ll 1$, of an accelerating Kerr spacetime does not coincide with the Kerr solution.


\begin{thebibliography}{99}


\bibitem{Kerr:1963ud} 
R.~P.~Kerr,
Phys.\ Rev.\ Lett.\  {\bf 11}, 237 (1963).

\bibitem{Hawking:1971vc} 
S.~W.~Hawking,
Commun.\ Math.\ Phys.\  {\bf 25}, 152 (1972).

\bibitem{Blandford:1977ds} 
R.~D.~Blandford and R.~L.~Znajek,
Mon.\ Not.\ Roy.\ Astron.\ Soc.\  {\bf 179}, 433 (1977).

\bibitem{McClintock:2006xd} 
J.~E.~McClintock, R.~Shafee, R.~Narayan, R.~A.~Remillard, S.~W.~Davis and L.~X.~Li,
Astrophys.\ J.\  {\bf 652}, 518 (2006).

\bibitem{Robinson:1975bv} 
D.~C.~Robinson,
Phys.\ Rev.\ Lett.\  {\bf 34}, 905 (1975).

\bibitem{Teukolsky:1973ha} 
S.~A.~Teukolsky,
Astrophys.\ J.\  {\bf 185}, 635 (1973).

\bibitem{Maldacena:1997re} 
J.~M.~Maldacena,
Int.\ J.\ Theor.\ Phys.\  {\bf 38}, 1113 (1999)
[Adv.\ Theor.\ Math.\ Phys.\  {\bf 2}, 231 (1998)].

\bibitem{Gubser:1998bc} 
S.~S.~Gubser, I.~R.~Klebanov and A.~M.~Polyakov,
Phys.\ Lett.\ B {\bf 428}, 105 (1998).

\bibitem{Witten:1998qj} 
E.~Witten,
Adv.\ Theor.\ Math.\ Phys.\  {\bf 2}, 253 (1998)

\bibitem{Aharony:1999ti} 
O.~Aharony, S.~S.~Gubser, J.~M.~Maldacena, H.~Ooguri and Y.~Oz,
Phys.\ Rept.\  {\bf 323}, 183 (2000).

\bibitem{Guica:2008mu} 
M.~Guica, T.~Hartman, W.~Song and A.~Strominger,
Phys.\ Rev.\ D {\bf 80}, 124008 (2009).

\bibitem{Brown:1986nw} 
J.~D.~Brown and M.~Henneaux,
Commun.\ Math.\ Phys.\  {\bf 104}, 207 (1986).


\bibitem{Carlip:2011ax} 
S.~Carlip,
JHEP {\bf 1104}, 076 (2011).

\bibitem{Compere:2012jk} 
G.~Compère,
Living Rev.\ Rel.\  {\bf 15}, 11 (2012)
[Living Rev.\ Rel.\  {\bf 20}, no. 1, 1 (2017)].

\bibitem{Siahaan:2015xia} 
H.~M.~Siahaan,
Class.\ Quant.\ Grav.\  {\bf 33}, no. 15, 155013 (2016).

\bibitem{Castro:2010fd} 
A.~Castro, A.~Maloney and A.~Strominger,
Phys.\ Rev.\ D {\bf 82}, 024008 (2010).

\bibitem{Chen:2010ik} 
B.~Chen and J.~Long,
Phys.\ Rev.\ D {\bf 82}, 126013 (2010).

\bibitem{Ghezelbash:2012qn} 
A.~M.~Ghezelbash and H.~M.~Siahaan,
Class.\ Quant.\ Grav.\  {\bf 30}, 135005 (2013).

\bibitem{Chen:2010as} 
C.~M.~Chen and J.~R.~Sun,
JHEP {\bf 1008}, 034 (2010)

\bibitem{Chen:2010fr} 
B.~Chen, J.~Long and J.~j.~Zhang,
Phys.\ Rev.\ D {\bf 82}, 104017 (2010).

\bibitem{Chen:2010ywa} 
C.~M.~Chen, Y.~M.~Huang, J.~R.~Sun, M.~F.~Wu and S.~J.~Zou,
Phys.\ Rev.\ D {\bf 82}, 066004 (2010).

\bibitem{Chen:2011kt} 
B.~Chen and J.~j.~Zhang,
JHEP {\bf 1108}, 114 (2011).

\bibitem{Majhi:2015tpa} 
B.~R.~Majhi,
Phys.\ Rev.\ D {\bf 92}, no. 6, 064026 (2015).

\bibitem{Franzin:2011wi} 
E.~Franzin and I.~Smolic,
JHEP {\bf 1109}, 081 (2011).

\bibitem{Shi:2018lpw} 
C.~Shi, J.~D.~Zhang and J.~Mei,
JHEP {\bf 1804}, 001 (2018).

\bibitem{Kinnersley:1970zw} 
W.~Kinnersley and M.~Walker,
Phys.\ Rev.\ D {\bf 2}, 1359 (1970).


\bibitem{Dias:2002mi} 
O.~J.~C.~Dias and J.~P.~S.~Lemos,
Phys.\ Rev.\ D {\bf 67}, 064001 (2003).

\bibitem{Plebanski:1976gy} 
J.~F.~Plebanski and M.~Demianski,
Annals Phys.\  {\bf 98}, 98 (1976).

\bibitem{Griffiths:2005qp} 
J.~B.~Griffiths and J.~Podolsky,
Int.\ J.\ Mod.\ Phys.\ D {\bf 15}, 335 (2006).

\bibitem{Griffiths:2009dfa} 
J.~B.~Griffiths and J.~Podolsky,
Cambridge University Press (2009).

\bibitem{Hong:2003gx} 
K.~Hong and E.~Teo,
Class.\ Quant.\ Grav.\  {\bf 20}, 3269 (2003).

\bibitem{Hong:2004dm} 
K.~Hong and E.~Teo,
Class.\ Quant.\ Grav.\  {\bf 22}, 109 (2005).

\bibitem{Astorino:2016xiy} 
M.~Astorino,
Phys.\ Lett.\ B {\bf 760}, 393 (2016).

\bibitem{Appels:2016uha} 
M.~Appels, R.~Gregory and D.~Kubiznak,
Phys.\ Rev.\ Lett.\  {\bf 117}, no. 13, 131303 (2016).

\bibitem{Astorino:2016ybm} 
M.~Astorino,
Phys.\ Rev.\ D {\bf 95}, no. 6, 064007 (2017).


\bibitem{Appels:2017xoe} 
M.~Appels, R.~Gregory and D.~Kubiznak,
JHEP {\bf 1705}, 116 (2017).

\bibitem{Dudley:1977zz} 
A.~L.~Dudley and J.~D.~Finley,
Phys.\ Rev.\ Lett.\  {\bf 38}, 1505 (1977).

\bibitem{Dudley:1978vd} 
A.~L.~Dudley and J.~D.~Finley, III,
J.\ Math.\ Phys.\  {\bf 20}, 311 (1979).

\bibitem{Bini:2014kga} 
D.~Bini, D.~Bini, C.~Cherubini and A.~Geralico,
J.\ Math.\ Phys.\  {\bf 49}, 062502 (2008).

\bibitem{Bini:2015xpa} 
D.~Bini, E.~Bittencourt and A.~Geralico,
Class.\ Quant.\ Grav.\  {\bf 32}, no. 21, 215010 (2015)

\bibitem{Kofron:2015gli} 
D.~Kofroň,
Phys.\ Rev.\ D {\bf 92}, no. 12, 124064 (2015).

\bibitem{Kofron:2016dyk} 
D.~Kofroň,
Phys.\ Rev.\ D {\bf 93}, no. 10, 104012 (2016).


\bibitem{Chen:2011wm} 
B.~Chen and J.~j.~Zhang,
Nucl.\ Phys.\ B {\bf 856}, 449 (2012).

\bibitem{Bredberg:2011hp} 
I.~Bredberg, C.~Keeler, V.~Lysov and A.~Strominger,
Nucl.\ Phys.\ Proc.\ Suppl.\  {\bf 216}, 194 (2011).

\bibitem{Wang:2010qv} 
Y.~Q.~Wang and Y.~X.~Liu,
JHEP {\bf 1008}, 087 (2010).

\bibitem{ChangYoung:2012kd} 
E.~Chang-Young and M.~Eune,
JHEP {\bf 1305}, 018 (2013).

\bibitem{Majhi:2014lka} 
B.~R.~Majhi,
Phys.\ Rev.\ D {\bf 90}, no. 4, 044020 (2014).

\bibitem{Ghezelbash:2013uga} 
A.~M.~Ghezelbash and H.~M.~Siahaan,
Phys.\ Rev.\ D {\bf 89}, no. 2, 024017 (2014).


\bibitem{Parikh:1999mf} 
M.~K.~Parikh and F.~Wilczek,
Phys.\ Rev.\ Lett.\  {\bf 85}, 5042 (2000).

\bibitem{Vanzo:2011wq} 
L.~Vanzo, G.~Acquaviva and R.~Di Criscienzo,
Class.\ Quant.\ Grav.\  {\bf 28}, 183001 (2011).

\bibitem{Angheben:2005rm} 
M.~Angheben, M.~Nadalini, L.~Vanzo and S.~Zerbini,
JHEP {\bf 0505}, 014 (2005).

\bibitem{Banerjee:2008cf} 
R.~Banerjee and B.~R.~Majhi,
JHEP {\bf 0806}, 095 (2008).


\bibitem{Siahaan:2009qv} 
H.~M.~Siahaan and Triyanta,
Int.\ J.\ Mod.\ Phys.\ A {\bf 25}, 145 (2010).

\bibitem{Henry:1999rm} 
R.~C.~Henry,
Astrophys.\ J.\  {\bf 535}, 350 (2000).

\end{thebibliography}
\end{document}